# Low-Energy, Octave-Spanning Supercontinuum Generation in $Ta_2O_5$ Waveguides: Towards Optical Coherence Metrology

Yanyan Zhang[1], Qin Xie[2], Xiaoqing Chen[1]*, Xuetao Gan[3]*.

1. School of Artificial Intelligence, OPtics and ElectroNics, Northwestern Polytechnical University, Xi'an, China. 710072)

2. School of Integrated Circuits, Northwestern Polytechnical University, Xi'an, China. 710129)

3. Key Laboratory of Light Field Manipulation and Information Acquisition, Ministry of Industry and Information Technology, and Shanxi Key Laboratory of Optical Information Technology, School of Physical Science and Technology, Northwestern Polytechnical University, Xi'an, China. 710129)

**Email:**

Yanyan Zhang[1], zhangyanyan@nwpu.edu.cn;

Qin Xie[2], 2024204570@mail.nwpu.edu.cn;

Xiaoqing Chen[1]*, xiaoqing_chen@nwpu.edu.cn;

Xuetao Gan[3]*, xuetaogan@nwpu.edu.cn;

## Abstract

Supercontinuum generation (SCG) on integrated photonic platforms is a pivotal technology for developing next-generation chip-scale systems for precision spectroscopy and metrology. While significant progress has been made with silicon (Si) and silicon nitride ($Si_3N_4$) platforms, they are often constrained by two-photon absorption(TPA) or moderate nonlinear coefficients, necessitating a trade-off between energy efficiency and bandwidth. Tantalum pentoxide ($Ta_2O_5$), possessing both high nonlinearity and a wide bandgap, emerges as a promising candidate; however, current implementations remain challenged by high pump energy consumption. Here, we report a low-loss $Ta_2O_5$ integrated waveguide fabricated via the Damascene process. It enables the generation of a two-octave-spanning spectrum with a low pulse energy of only 92.9 pJ (60 fs, 1550 nm). Notably, the corresponding peak power is a mere 1.36 kW, which is nearly an order of magnitude lower than that of state-of-the-art comparable broadband sources. Furthermore, at the maximum pump energy, our spectrum exhibits an ultrabroad coverage from 450 nm to 3400 nm, spanning nearly three octaves. Supported by numerical simulations, we analyze the dynamics of soliton fission. Furthermore, a Michelson interferometry system developed using this source exhibits superior performance, achieving not only micrometer-scale axial resolution but also a 6 dB sensitivity roll-off length of 3.1 mm. This exceptional roll-off performance, combined with a displacement measurement sensitivity of 346 nm, underscores the immense potential of the $Ta_2O_5$ platform for applications in biomedical imaging and precision metrology.

## Introduction

With the rapid evolution of precision manufacturing and quantum sensing technologies, optical metrology is undergoing a critical transformation from benchtop laboratory equipment to chip-scale integrated systems[1-4]. Central to this transition is the development of compact, high-brightness coherent light sources with ultra-broad spectral coverage, which has emerged as a key enabler for breaking through existing detection limits. SCG in integrated photonic waveguides exploits dispersion engineering at the micro- and nanoscale to stimulate cascaded nonlinear dynamic processes, thereby efficiently converting narrowband laser light into broadband spectra spanning over an octave[5,6]. Merging the high spatial coherence of lasers with the ultra-wide bandwidth characteristic of conventional light sources, these devices offer an ideal solution for next-generation chip-scale precision spectroscopy, optical frequency combs, and coherent interferometry[7-10].

Although Si and $Si_3N_4$ have established themselves as universal platforms for on-chip nonlinear optics due to their mature CMOS-compatible fabrication processes, they face an inherent trade-off between performance and energy efficiency when pursuing optimal power budgets. Si is severely constrained by significant TPA in the telecommunications band[11], which clamps its nonlinear conversion efficiency at high power densities. Conversely, while $Si_3N_4$ exhibits extremely low linear loss and is free from TPA, its moderate nonlinear refractive index[12] ($n_2 \approx 2.4 \times 10^{-15}$ cm²/W) often necessitates the use of centimeter-scale waveguides or excessively high pump powers to accumulate sufficient nonlinear phase shift. Such reliance on extended interaction lengths or elevated pump powers inevitably escalates system packaging complexity and energy consumption budgets.

Against this backdrop, $Ta_2O_5$ emerges as a uniquely competitive candidate due to its combination of a wide bandgap and high nonlinearity. It boasts an ultra-wide transparency window spanning from the ultraviolet to the mid-infrared, alongside a Kerr nonlinear coefficient approximately three times higher than that of $Si_3N_4$[13] ($n_2 \approx$ 4-7 $\times 10^{-15}$ cm²/W). This elevated nonlinearity facilitates efficient spectral broadening within short, millimeter-scale waveguides. However, a review of existing research on $Ta_2O_5$ SCG[14-19] reveals that device performance remains constrained by an inherent trade-off between spectral bandwidth and pump energy consumption. While some studies have demonstrated SCG, their spectral coverage typically falls short of 1.6 octaves[15,16]; conversely, achieving ultra-broadband output spanning multiple octaves has required pulse energies approaching the nanojoule (nJ) level[17]. This inefficient paradigm of "trading high energy consumption for broad bandwidth" largely negates the low-power advantages promised by integrated photonics, thereby restricting the platform's practical utility in portable precision measurement systems.

To overcome this energy efficiency bottleneck, we propose a $Ta_2O_5$ integrated waveguide scheme fabricated via the Damascene process. By employing a "trench-first, fill-later" strategy, this technique significantly mitigates scattering losses induced by sidewall roughness, thereby ensuring the low-loss propagation of ultrashort pulses over extended distances within high-confinement modes. When combined with our superior dispersion engineering, we achieved a highly competitive, low-threshold SCG: experimental results demonstrate that a coupled pulse energy of merely sub-hundred picojoules is sufficient to excite an ultra-broadband spectrum extending from the visible to the mid-infrared regime within the waveguide.

Building on this foundation, we further conducted a systematic evaluation of the coherence properties and metrological capabilities of the source—an aspect rarely addressed in comparable studies. We established a Michelson interferometer-based system to perform in-depth quantitative characterization of the point spread function (PSF) and sensitivity roll-off dynamics under broadband operation. The measurement results reveal that, underpinned by excellent spectral coherence and ultra-wide bandwidth, the system not only achieves micrometer-scale axial resolution but also extends the sensitivity roll-off length to the 3.1 mm level. Furthermore, step tests using a nano-displacement stage confirmed the system's exceptional measurement linearity, with a minimum resolvable displacement (MRD) as low as 346 nm. This comprehensive verification of key physical parameters vividly demonstrates the immense potential of this $Ta_2O_5$ source as a core illumination platform for next-generation high-resolution biomedical imaging and deep-tissue non-destructive sensing.

## Results

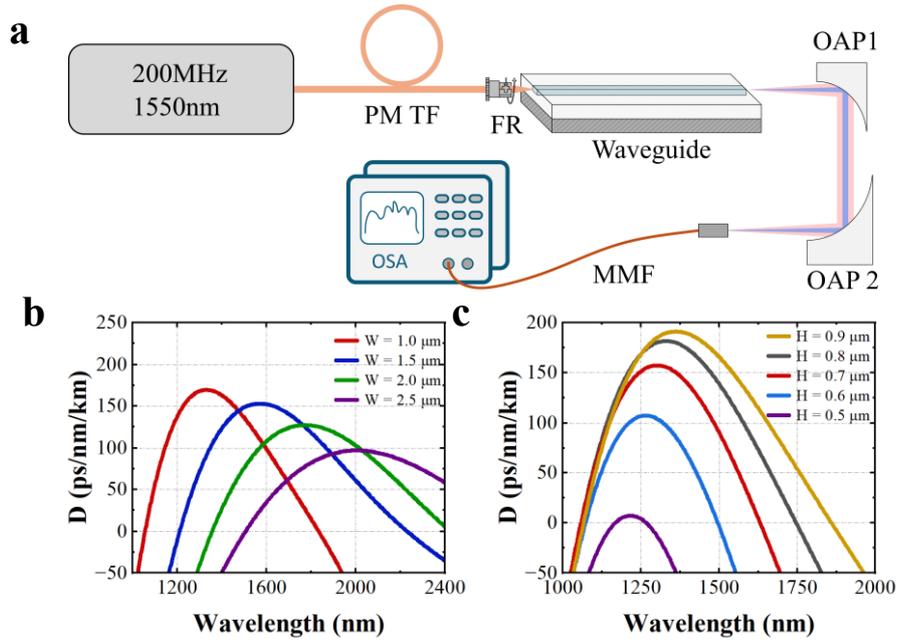

**Fig. 1 Ta$_2$O$_5$ waveguide platform: experimental setup, dispersion characteristics, and mode field distribution. a** Schematic of the experimental setup for SCG. The seed source is a 200MHz, 1550nm femtosecond laser. PM TF: polarization-maintaining tapered fiber; FR: fiber rotator (for input polarization adjustment); OAP: off-axis parabolic mirror; MMF: multimode fluoride fiber; OSA: optical spectrum analyzer. **b** Calculated group velocity dispersion (GVD) curves for waveguide widths varying from 1.0 µm to 2.5 µm (with a fixed height of 0.8 µm). **c** GVD curves for waveguide heights varying from 0.5 µm to 0.9 µm (with a fixed width of 1.0 µm).

The experimental setup for SCG is illustrated in Fig. 1a. To optimize spectral broadening performance, we first numerically investigated the dependence of dispersion characteristics on waveguide geometry using the finite element method[20]. As depicted in Figs. 1b and 1c, the GVD curves exhibit a strong dependence on variations in waveguide width and height. In nonlinear optics, anomalous dispersion is a prerequisite for triggering soliton dynamics and achieving broadband spectral generation. After comprehensive consideration, we selected a waveguide cross-section with dimensions of $W$=1.0 μm and $H$=800 nm. This geometric configuration provides substantial anomalous dispersion at the 1550 nm pump wavelength, thereby establishing robust phase-matching conditions for efficient soliton fission and dispersive wave radiation[21].

High-quality waveguide fabrication is the foundation for achieving low-loss transmission and high-performance on-chip photonic emission. In contrast to traditional subtractive manufacturing, this work employs the Damascene process to fabricate $Ta_2O_5$ waveguides, thereby minimizing scattering losses induced by sidewall roughness. The fabrication process commences with a Si substrate featuring a 10 μm-thick thermal silicon dioxide layer, which serves as the bottom cladding. First, waveguide trench patterns were precisely defined using electron beam lithography and transferred into the $SiO_2$ layer via inductively coupled plasma etching. To further suppress scattering, a critical thermal reflow treatment was performed post-etching to smooth the trench sidewalls[22]. Subsequently, a high-quality $Ta_2O_5$ film was deposited via magnetron sputtering to fully fill the trenches. The specific process conditions were as follows: a $Ta_2O_5$ target was used under an RF power of 100 W, a substrate temperature of 200 °C, and a chamber pressure of 0.3 Pa (with $Ar/O_2$ flow rates of 20/5 sccm), resulting in a deposition rate of approximately 6.06 nm/min. Finally, chemical mechanical polishing was employed to remove excess material from the top surface and achieve global planarization of the waveguide. Details of the complete Damascene fabrication process, along with the actual cross-sectional morphology and the corresponding eigenmode profile, are provided in Supplementary Fig. S1. The SEM and mode simulation results verify that the Damascene process endows the waveguide with excellent core shaping quality and ultralow sidewall roughness, while ensuring tight confinement of the fundamental mode within the $Ta_2O_5$ core. Such strong optical confinement significantly boosts the nonlinear parameter ($\gamma$), maximizing the light-matter interaction.

We further characterized the linear optical performance of the device by measuring its total insertion loss. Experimentally, the total insertion loss of the fiber-chip-fiber link was measured to be below 14 dB. This characterization of linear transmission properties provides a reliable energy benchmark for subsequent nonlinear experiments, ensuring that femtosecond pump pulses maintain extremely high peak power densities over sufficient interaction lengths to effectively surpass the excitation threshold for nonlinear effects.

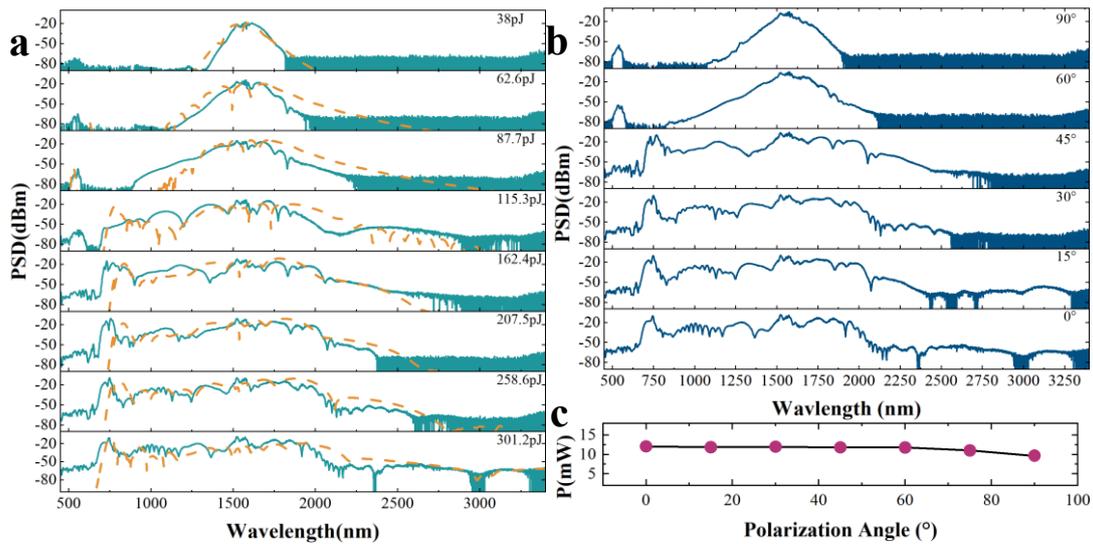

**Fig. 2 Experimental characterization of SCG in $Ta_2O_5$ waveguides. a** Spectral evolution as the coupled pulse energy increases from 38 pJ to 301.2 pJ. Solid lines (cyan) represent experimental

measurements, while dashed lines (orange) denote numerical simulations based on the generalized nonlinear Schrödinger equation (GNLSE). The input pulse duration is 80 fs with a central wavelength of 1550 nm. **b** Evolution of the supercontinuum spectra with the incident polarization angle varying from 0° (TE mode) to 90° (TM mode) under maximum pumping conditions. **c** Corresponding average output power as a function of the polarization angle.

To systematically characterize the SCG performance of the $Ta_2O_5$ waveguides, we first conducted experimental verification using pulses with a duration of 80 fs. Fig. 2a illustrates the spectral evolution as the coupled pulse energy increases from 38 pJ to 301.2 pJ, clearly revealing the dynamic transition from a self-phase modulation-dominated regime to intense soliton fission.

In the low-energy regime (38 pJ-62.6 pJ), the spectra primarily exhibit symmetric self-phase modulation oscillatory structures centered at 1550 nm, lacking distinct redshifted features. This is attributed to an insufficient effective evolution distance. Based on soliton theory calculations, the soliton order ($N$) at this stage reaches 3.8-4.9, corresponding to a fission length ($L_{fiss}$) of approximately 2.9-3.7 mm. Considering the total waveguide length of 6.8 mm, soliton fission occurs only in the latter half of the waveguide. Consequently, the fundamental solitons ejected from the fission process have less than 3 mm of propagation distance remaining, preventing them from accumulating sufficient Raman self-frequency shift to achieve significant spectral separation from the pump in the frequency domain. However, as the pump energy is further increased, the dynamic mechanism undergoes a fundamental transformation.

The soliton fission point shifts dramatically to the vicinity of the waveguide input (approximately 1.3 mm, as indicated in Fig. S2a). This affords an ample evolution distance of over 5 mm (>80% of the total waveguide length) for nonlinear processes, allowing high-energy fundamental solitons to continuously redshift under the drive of the strong Raman effect, ultimately forming a flat long-wave tail extending beyond 3000 nm.

Notably, the experimentally measured spectral envelopes (solid lines) exhibit a high degree of consistency with simulation results based on the GNLSE (dashed lines) across all stages, underscoring the high optical quality of the fabricated waveguides and the accuracy of the theoretical model.

The polarization state of the input light represents another critical degree of freedom for tailoring nonlinear interactions in integrated waveguides. We further investigated the dependence of SC generation on the incident polarization state. As shown in Fig 2b, by rotating the polarization direction of the incident light, we recorded the spectral response varying from the TE mode (0°) to the TM mode (90°). The experimental results reveal that the TE mode exhibits significantly broader spectral coverage, whereas spectral broadening in the TM mode is relatively constrained.

To elucidate the microscopic physical mechanisms underlying this mode dependence, we combined full-vector finite element method simulations with GNLSE analysis (detailed in Supplementary Note 2 & 3). The simulation results suggest that the advantage of the TE mode stems primarily from two factors: First, the TE mode possesses a flatter dispersion profile and a smaller effective mode area (Fig. S2c) compared to the TM mode—which is prone to premature "pulse collapse"—thereby favoring the effective accumulation of nonlinear phase shift and spectral broadening. Second, the TE mode field remains continuous at the waveguide boundaries, mitigating scattering losses triggered by the interface field enhancement effect observed in the TM mode[23] (Fig. S3). Despite this, the device exhibits exceptional polarization robustness in the mid-infrared regime across the full angular range. Coupled with its low output power fluctuation (<20%) (Fig. 2c), this source provides a solid foundation for polarization-insensitive infrared sensing applications.

To further explore the bandwidth limits of this $Ta_2O_5$ platform, we attempted to compress the pump pulse to 60 fs. Under this extreme driving condition, we observed an ultra-broad supercontinuum spanning from the visible to the mid-infrared (0.45-3.4 μm) (Supplementary Fig. S4), signifying the platform's immense potential for bandwidth extension.

Given the superior nonlinear performance described above, we proceeded to systematically characterize the coherence properties of the source and its performance in interferometry under typical 80 fs operating conditions, aiming to evaluate its utility in precision measurement scenarios.

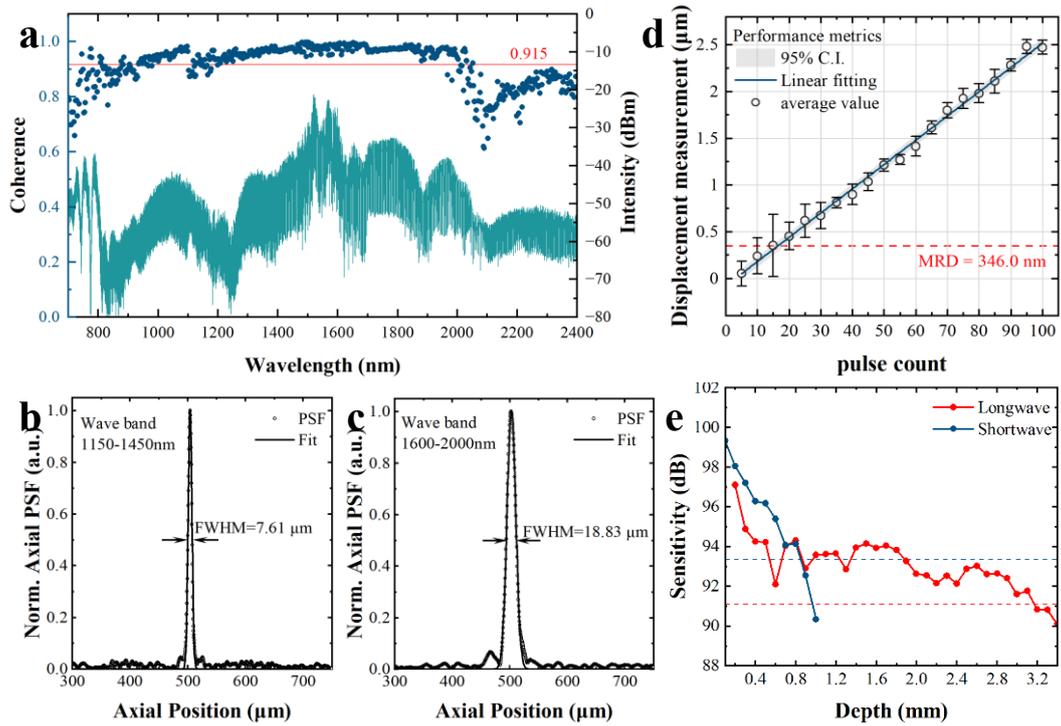

Fig. 3 Characterization of source coherence and evaluation of precision measurement performance. a First-order spectral degree of coherence ($|g_{12}^{(1)}|$, blue dots, left axis) measured over the broad spectral range of 700-2400 nm, plotted alongside the corresponding spectral intensity (cyan curve, right axis). The red line indicates the mean coherence (0.915). b, c Normalized axial PSF retrieved using the source in the short-wavelength band (1150-1450 nm) (b) and long-wavelength band (1600-2000 nm) (c). Gaussian fits (solid lines) yield full-width at half-maxima of 7.61 μm and 18.83 μm, respectively. d Results of nano-displacement measurement. The measured displacement exhibits a highly linear relationship with the stage step number ($R^2 = 0.996$). The MRD, determined based on the 3σ criterion, is 346 nm. The gray band represents the 95% confidence interval of the fit. e Sensitivity roll-off curves for the short-wavelength (blue) and long-wavelength (red) bands. The 6 dB roll-off lengths reach 0.8 mm and 3.1 mm for the short and long bands, respectively.

To comprehensively evaluate the spatiotemporal characteristics of this supercontinuum source and its potential for applications in precision measurement, we first conducted a quantitative characterization of the spectral coherence. Figure 3a presents the modulus of the first-order complex degree of coherence[24-26] measured over the 700 nm to 2400 nm spectral band. Although the spectral intensity (cyan curve) exhibits certain energy fluctuations across different bands, our source demonstrates a high average coherence of approximately 0.915. This high coherence ensures the generation of high-contrast interference fringes in broadband interferometry, thereby establishing a solid physical foundation for high-signal-to-noise ratio signal demodulation[27].

Based on the superior ultra-broadband coherence characteristics of the source, we constructed a Michelson interferometer system to evaluate its performance limits in precision measurement (detailed in Supplementary Fig. S5, which includes the experimental setup and raw interference signals). The ultra-wide spectral coverage grants us the capability to flexibly tune measurement characteristics using a "spectral windowing" strategy. At a fixed optical path difference, we examined the interference signals in the short-wavelength telecommunication band (1150-1450 nm) and the mid-infrared sensing band (1600-2000 nm), respectively, and retrieved their axial PSF.

As shown in Fig 3b, the PSF at short wavelengths exhibits an extremely narrow main lobe, achieving a high axial resolution of approximately 7.61 μm. In contrast, in the long-wavelength band (1600-2000 nm), due to the redshift of the central wavelength and variations in the spectral envelope, the corresponding resolution was measured to be approximately 18.83 μm (Fig 3c). This wavelength-dependent resolution characteristic reveals a tunable trade-off mechanism: the short-wavelength window can be prioritized when fine structural analysis is required, while the long-wavelength window offers superior penetration depth when facing highly scattering biological tissues or materials. To validate this, we measured the sensitivity roll-off characteristics of the system (Fig 3e). The results indicate that the 6 dB roll-off length is 0.8 mm in the short-wavelength band, whereas it reaches 3.1 mm in the long-wavelength band, underscoring its unique advantages in deep structural detection[10,28].

Finally, to rigorously define the measurement repeatability and detection limits of the system, we performed a series of step displacement tests using a nano-displacement stage (Fig S6). As shown in the upper panel of Figure 3d, the measured displacement exhibits a highly linear relationship with the stage pulse number ($R^2 = 0.996$), demonstrating the accuracy of the system's quantitative measurements. When assessing the MRD, we adopted the strict $3\sigma$ criterion[29] to ensure sufficient statistical confidence in the results. Through statistical analysis of the residual distribution from multiple repeated measurements, the displacement sensitivity of the system was calculated to be 346 nm. This sub-micron sensitivity metric implies that the system can effectively distinguish minute displacement changes within a 99.7% confidence interval, showcasing the application potential of this light source in the field of precision metrology.

|  | Pump (μm) | Pulse width (fs) | Repetition Frequency (MHz) | Spanning (octave) | Coverage (μm) | Launching Pulse Energy (pJ) | Peak Power(kW) | Reference |
|---|---|---|---|---|---|---|---|---|
| **Si₃N₄** | 1.55 | 90 | 100 | 2.68 | 0.56-3.60 | 600 | 6.3 | 30 |
|  | 2.09 | 78 | 19 | 1.3 | 1.70-4.20 | 750 | 6.8 | 31 |
|  | 1.56 | 120 | 40 | 2.2 | 0.53-2.60 | 1400 | 11.0 | 32 |
|  | 2.35 | 45 | 75 | 1.5 | 1.20-3.70 | 560 | 12.4 | 33 |
|  | 1.56 | 50 | 250 | 1 | 1.01-2.02 | 328 | 6.2 | 34 |
| **Si** | 1.95 | 250 | 80 | 1.26 | 1.00-2.40 | 45 | 0.15 | 35 |
|  | 4 | 300 | 0.15 | 1.32 | 2.00-5.00 | 28600 | 96 | 36 |
|  | 3.7 | 320 | 20 | 1.53 | 1.90-6.00 | 800 | 2.35 | 37 |
|  | 2.26 | 70 | 82 | 1.33 | 1.10-2.76 | 3.12 | 0.042 | 38 |
|  | 3.06 | 85 | 100 | 1.94 | 2.00-7.70 | 100 | 1.1 | 39 |
| **Ta₂O₅** | 1.056 | 100 | 80 | 1.5 | 0.59-1.70 | 37.5 | 0.35 | 15 |
|  | 1.56 | 80 | 100 | 1.6 | 0.78-2.40 | 900 | 10.6 | 14 |
|  | 1.56 | 80 | 100 | 1.6 | 0.75-2.40 | 60 | 0.71 | 13 |
|  | 1 | 200 | 80 | 1 | 0.70-1.48 | 2190 | 10.3 | 16 |
|  | 1.55&1.97 | 180&246 | 50 | 2.5 | 0.50-2.90 | 2130 | 7.9 | 17 |
|  | **1.55** | **60** | **200** | **2.52/2.92** | **0.45-3.40** | **92.9/299.79** | **1.36/4.40** | **This work** |

**Table 1 Comparison of on-chip SCG performance across different material platforms.** This table summarizes key parameters of representative integrated nonlinear waveguide light sources based on Si₃N₄, Si, and Ta₂O₅. The comparison metrics cover pump conditions (center wavelength, pulse duration, repetition rate) and output spectral characteristics (octave span, spectral coverage, coupled pulse energy, and peak power). Data from "This work" are highlighted in red, underscoring the achievement of a two-octave spanning spectrum with a low coupled pulse energy of only 92.9 pJ, driven by 60 fs ultrashort pulses at a high repetition rate of 200 MHz. Furthermore, a maximum ultra-broadband spectral coverage of 0.45-3.4 μm is realized, demonstrating the superior performance of our platform.

To rigorously evaluate the positioning of this work within the field of integrated photonics, Table 1 systematically summarizes the key performance metrics of classic supercontinuum sources based on Si, $Si_3N_4$, and $Ta_2O_5$ platforms.

Although Si and $Si_3N_4$ platforms have established mature fabrication ecosystems, they remain constrained by inherent physical bottlenecks when pursuing efficient ultra-broadband generation. Silicon is fundamentally limited in its nonlinear conversion efficiency at high power densities due to strong TPA in the telecommunications band. Conversely, while $Si_3N_4$ exhibits extremely low propagation loss, its moderate nonlinear coefficient often compels devices to rely on centimeter-scale interaction lengths or kilowatt-level peak powers to achieve spectral broadening. This dependence on extended physical dimensions and high-energy pumping largely negates the inherent advantages of compactness and low power consumption that integrated photonic devices are intended to offer.

In sharp contrast, this work highlights the unique competitiveness of the high-quality $Ta_2O_5$ platform fabricated via the Damascene process. As evidenced by the data in the table, in the competition among comparable $Ta_2O_5$ devices, prior high-performance demonstrations typically required pulse energies at the nanojoule (nJ) level (e.g., 2190 pJ[16,17] or 900 pJ[14]) to initiate octave-spanning nonlinear processes. This work successfully breaks this traditional paradigm of "trading high energy consumption for broad bandwidth," achieving an order-of-magnitude reduction in the energy threshold. Benefiting from the superior loss control and dispersion engineering afforded by the Damascene process, we achieved a two-octave spectral output with a coupled pulse energy of only 92.9 pJ (peak power 1.36 kW). Furthermore, at maximum capacity, we demonstrated an ultra-broadband spectral coverage spanning from 0.45 μm to 3.4 μm (a span of 2.92 octaves).

To the best of our knowledge, this result establishes a new record for the widest SCG in $Ta_2O_5$ waveguides to date. This breakthrough compellingly validates the platform's exceptional energy-to-bandwidth conversion efficiency. The ultimate balance achieved here between low power consumption, ultra-wide bandwidth, and compact footprint establishes this $Ta_2O_5$ platform as an ideal physical foundation for next-generation precision frequency metrology and spectral analysis systems.

## Discussion

In this study, we addressed the long-standing bottleneck of high energy consumption in the $Ta_2O_5$ platform by employing the Damascene process. Compared to existing literature, our devices exhibit superior nonlinear conversion efficiency: the onset of SCG spanning two octaves is triggered by a coupled pulse energy of merely 92.9 pJ—an order of magnitude lower than that of comparable devices fabricated via traditional subtractive manufacturing. This low-energy operation mode makes the use of compact, high-repetition-rate lasers as pump sources feasible. Furthermore, supported by the platform's exceptional power handling capabilities, we successfully extended the spectral boundary to 0.45-3.4 μm by further increasing the pump energy. This ultra-wide bandwidth, covering the visible to mid-infrared regimes, provides a highly coherent, broad-coverage spectral resource for molecular fingerprint spectroscopy.

Through comparative theoretical and experimental analysis, we elucidated the competition mechanisms between TE and TM modes in ultra-broadband generation. The superior bandwidth and efficiency of the TE mode are primarily attributed to its flatter dispersion profile and smaller effective mode area, which ensure that soliton fission and subsequent dispersive wave radiation occur more efficiently, thereby explaining its superior spectral broadening capability at the level of microscopic dynamics. Despite these modal differences, the stable output of the device across all polarization angles demonstrates its robustness in practical applications.

Beyond SCG, this work compellingly demonstrates the utility of this source in precision metrology through interferometric experiments. Notably, we observed exceptional sensitivity roll-off characteristics, achieving a 6 dB roll-off length of 3.1 mm in the long-wavelength band. This parameter is a critical metric for depth-resolved imaging systems such as optical coherence tomography[40,41], implying that the source can maintain high signal-to-noise ratio signals over millimeter-scale depths. Combined with its micrometer-scale axial resolution and nanometer-scale displacement sensitivity (346 nm), the system exhibits a comprehensive measurement capability that balances high-resolution imaging with deep structural penetration.

In summary, we have demonstrated a highly energy-efficient, ultra-broadband nonlinear photonic platform based on Damascene $Ta_2O_5$ waveguides. This platform successfully breaks the trade-off between energy consumption and bandwidth, realizing a spectral output covering nearly three octaves using picojoule-level pulses. Its capability for low-threshold octave generation, combined with superior interferometric roll-off performance, establishes the $Ta_2O_5$ waveguide as a pivotal component in next-generation low-power, portable optical diagnostics and precision metrology devices.

## Materials and methods

### Experimental Setup

The experiment utilized a mode-locked Erbium-doped fiber laser as the pump source, featuring a central wavelength of 1550 nm, a repetition rate of 200 MHz, and an initial pulse duration of 80 fs. After the polarization state was adjusted via a fiber polarization controller, the pump light was coupled into the $Ta_2O_5$ waveguide using a tapered lensed fiber with a mode field diameter of 3.5 μm. The SCG at the waveguide output was collimated by two OAPs and subsequently coupled into a multimode fluoride fiber. Spectral data were acquired using multiple optical spectrum analyzers (Yokogawa AQ6374E and AQ6376) to cover a broad spectral range from 0.45-3.4 μm.

### Numerical Simulation

The nonlinear propagation dynamics of ultrashort pulses within the waveguide were simulated by solving the GNLSE using the split-step Fourier method. The complex refractive index and frequency-dependent GVD profiles of the waveguide were calculated using a finite-element mode solver (COMSOL Multiphysics), with high-order dispersion terms retained in the simulation to accurately describe the soliton fission process. Additionally, the simulation comprehensively incorporated the Raman response function and self-steepening effect of $Ta_2O_5$. The nonlinear refractive index

($n_2$) of the material was set to $4.7 \times 10^{-15}$ cm²/W, consistent with the characteristics of the experimental material. The input pulse was modeled as a hyperbolic secant (Sech²) pulse, with parameters for pulse duration, central wavelength, and repetition rate kept consistent with experimental conditions. Detailed simulation parameters are provided in Supplementary Table S1.

**System Characterization and Statistical Analysis**

The theoretical axial resolution of the system is based on the formula:

$$\Delta z \approx \frac{2 \ln 2}{\pi} \frac{\lambda^2}{\Delta \lambda}$$

where $\lambda_c$ is the central wavelength and $\Delta\lambda$ is the full width at half maximum of the spectrum. To rigorously evaluate the displacement measurement precision of the system, we introduced the statistical $3\sigma$ criterion. Multiple sets of interference signals were acquired under static conditions at different system positions to calculate the average standard deviation of the retrieved measurement results. The MRD of the system is defined as $3\sigma$, representing the measurement limit within a 99.7% confidence interval.

**Sensitivity and Roll-off Characterization**

To quantitatively assess system performance, we characterized the sensitivity and its roll-off behavior at various imaging depths (Fig. S5). A mirror was placed in the sample arm as the test target. Measurements were performed by axially translating the OAP assembly in the reference arm using a precision displacement stage to vary the optical path difference. At each position, sensitivity was calculated as the signal-to-noise ratio Specifically, the signal power was determined from the peak magnitude of the PSF, while the noise floor was derived by calculating the standard deviation of the background. The 6 dB sensitivity roll-off is defined as the position at which the measured sensitivity drops by 6 dB relative to the maximum value at the zero-delay position.

## Acknowledgements

## Conflict of Interest

The authors declare no conflicts of interest.

# Supplementary Information for

# Low-Energy, Octave-Spanning Supercontinuum Generation in Ta$_2$O$_5$ Waveguides: Towards Optical Coherence Metrology

Yanyan Zhang[1], Qin Xie[2], Xiaoqing Chen[1]*, Xuetao Gan[3]*.

[1]School of Artificial Intelligence, OPtics and ElectroNics, Northwestern Polytechnical University, Xi'an, 710072, China.

[2]School of Integrated Circuits, Northwestern Polytechnical University, Xi'an, 710129, China.

[3]Key Laboratory of Light Field Manipulation and Information Acquisition, Ministry of Industry and Information Technology, and Shaanxi Key Laboratory of Optical Information Technology, School of Physical Science and Technology, Northwestern Polytechnical University, Xi'an 710129, China

*Email: xiaoqing_chen@nwpu.edu.cn; xuetaogan@nwpu.edu.cn.

## 1. Detailed Damascene Fabrication Process and Waveguide Characterization

The $Ta_2O_5$ waveguides in this work employ a "trench-first, fill-later" strategy based on the Damascene process. This approach fundamentally suppresses the linear scattering loss typically caused by sidewall roughness in conventional subtractive manufacturing. The complete fabrication workflow is illustrated in Fig. S1a and comprises the following core steps:

**Lithography and Etching:** The fabrication begins with a silicon substrate featuring a 10-μm-thick thermally oxidized $SiO_2$ layer, which serves as the lower cladding. Electron-beam (E-beam) lithography was used to precisely define the waveguide trench patterns, which were subsequently transferred into the $SiO_2$ layer via inductively coupled plasma (ICP) etching.

**Thermal Reflow:** To fully eliminate the sidewall roughness induced by the dry etching process, a critical thermal reflow treatment was applied to the etched $SiO_2$ trenches. This step effectively smooths out the etching damage.

**Material Deposition (Sputtering):** High-quality $Ta_2O_5$ thin films were deposited using radio-frequency (RF) magnetron sputtering to completely fill the waveguide trenches.

**Planarization:** Following the deposition, chemical mechanical polishing (CMP) was employed to remove the excess $Ta_2O_5$ material outside the trenches, ultimately achieving a globally planarized waveguide surface.

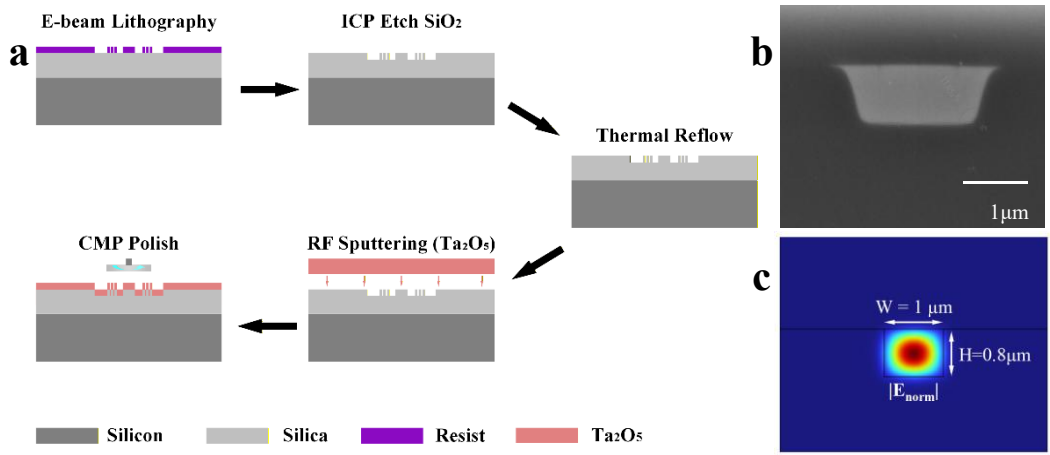

**Fig. S1. Damascene fabrication process and waveguide characterization. (a)** Schematic workflow of the Ta$_2$O$_5$ waveguide fabrication. The process sequentially includes electron-beam lithography patterning, SiO$_2$ dry etching (ICP), thermal reflow for sidewall smoothing, Ta$_2$O$_5$ deposition via RF sputtering, and final planarization using CMP. **(b)** Cross-sectional SEM image of a representative waveguide sample fabricated using the identical core process. The trench profile exhibits a calibrated sidewall angle of 108.5°, highlighting the ultra-low roughness achieved by the thermal reflow step. **(c)** Numerical simulation of the fundamental TE mode electric field distribution ($|E_{norm}|$) at a wavelength of 1550 nm, calculated based on the actual waveguide geometry.

To visually verify the improvement in waveguide quality enabled by this process, we performed scanning electron microscopy characterization on the device cross-section (Fig. S1b). It should be particularly noted that this scanning electron microscopy image is taken from a representative sample fabricated using the exact same core process flow (covering E-beam lithography, dry etching, thermal reflow smoothing, and magnetron sputtering deposition) as the actual tested devices. Due to considerations regarding cross-sectional cleaving and sample preparation, the top of this specific representative sample does not exhibit the final fully CMP-planarized state; however, this does not affect the accurate evaluation of the formation quality of the waveguide core region.

The core advantages of this process can be clearly observed from the image: benefiting from the crucial post-etching thermal reflow treatment, the waveguide trench exhibits an extremely smooth sidewall profile. Such a high-quality smooth interface serves as the physical foundation for achieving low-loss transmission. Furthermore, finite-element simulations based on the actual cross-sectional geometric parameters (Fig. S1c) indicate that the electric field of the fundamental TE mode at the central wavelength of 1550 nm is tightly and perfectly confined within the $Ta_2O_5$ core. This excellent mode localization characteristic significantly enhances nonlinear interactions, providing a crucial prerequisite for the subsequent efficient excitation of octave-spanning nonlinear dynamic evolution.

## 2. Dispersion Engineering and Nonlinear Propagation Dynamics

To thoroughly investigate the nonlinear propagation characteristics of ultrashort pulses in Tantalum pentoxide ($Ta_2O_5$) waveguides and validate the experimentally observed spectral evolution, we numerically solved the generalized nonlinear Schrödinger equation[1] (GNLSE) using the split-step Fourier method.

**(1) Theoretical Model**

The evolution of the pulse envelope $A(z,t)$ along the waveguide is governed by the following equation:

$$\frac{\partial A}{\partial z} + \frac{\alpha}{2}A - i\sum_{k\geq 2}\frac{i^k \beta_k}{k!}\frac{\partial^k A}{\partial t^k} = i\gamma\left(1 + \frac{i}{\omega_0}\frac{\partial}{\partial t}\right)\left(A(z,t)\int_{-\infty}^{\infty} R(t')|A(z,t-t')|^2\, dt'\right)$$

where:

$z$ is the propagation distance along the waveguide;

$t$ represents the time in a co-moving reference frame;

$\alpha(\omega)$ denotes the frequency-dependent linear loss profile;

$\beta_k$ are the $k$-th order dispersion coefficients, derived from the Taylor series expansion of the effective refractive index $n_{\text{eff}}(\omega)$ obtained via finite-element simulations (COMSOL).

The terms on the right-hand side of the equation describe the Kerr nonlinearity, the self-steepening effect (optical shock term), and the Raman response, respectively. The Raman response function is defined as $R(t) = (1-f_R)\delta(t) + f_R h_R(t)$, where $f_R$ represents the fractional contribution of the delayed Raman response[2].

**(2) Simulation Parameters**

The physical parameters adopted in our model are detailed in Supplementary Table

S1. Specifically, the nonlinear refractive index of $Ta_2O_5$ was set to $n_2 \approx 4\text{-}7 \times 10^{-15}$ cm²/W, and the fractional Raman contribution was set to $f_R = 0.18$. The input pulse was modeled as a hyperbolic secant (Sech²) waveform with a central wavelength of 1550 nm and a full-width at half-maximum of 80 fs, with the peak power kept consistent with experimental conditions.

Supplementary Table S1 Key physical parameters used in GNLSE simulations.

| Parameter | Symbol | Value | Unit |
| --- | --- | --- | --- |
| Nonlinear refractive index | $n_2$ | $4.7 \times 10^{-15}$ | cm²/W |
| Fractional Raman contribution | $f_R$ | 0.18 | - |
| Effective mode area | $A_{eff}$ | 0.78 (TE) / 0.85 (TM) | µm² |
| Pump center wavelength | $\lambda_p$ | 1550 | nm |
| Pulse width (FWHM) | $\tau$ | 80 | fs |
| Repetition rate | $f_{rep}$ | 200 | MHz |
| Peak power | $P_0$ | 4400 | W |

**(3) Dispersion Engineering and Spectral Evolution Mechanisms**

The dispersion profiles presented in Fig S2c provide the critical physical foundation for understanding the distinct spectral evolution observed in the two modes. At the pump wavelength of 1550 nm, although both the TE and TM fundamental modes lie within the anomalous dispersion regime, their dispersion behaviors differ significantly: the TM mode (red line) exhibits a large anomalous dispersion magnitude and a steep dispersion slope, whereas the TE mode (black line) possesses a relatively flatter dispersion profile with a smaller magnitude. This disparity directly dictates the characteristic interaction lengths of the pulse within the waveguide—specifically, the competition between the dispersion length $L_D$) and the nonlinear length ($L_{NL}$).

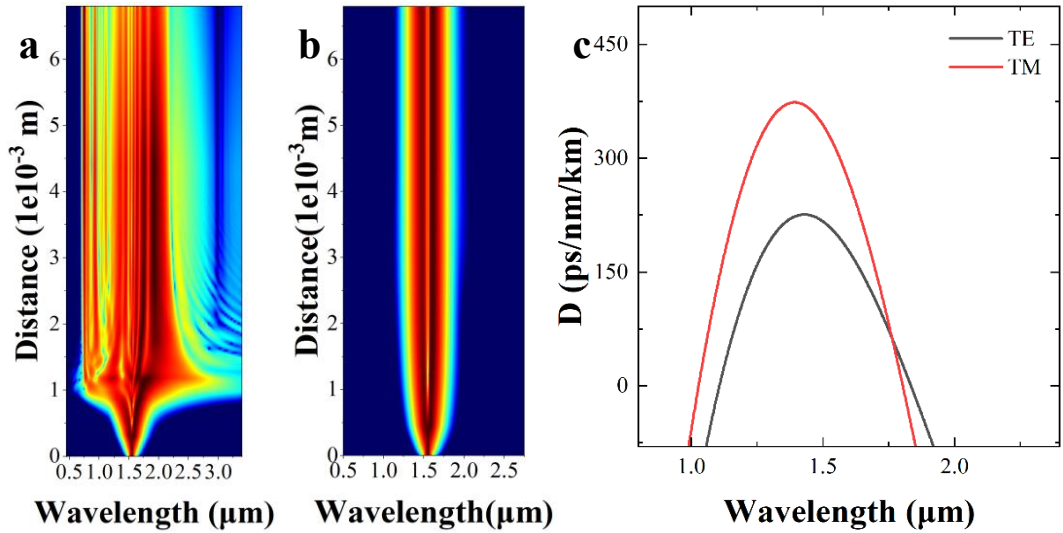

**Fig S2 Dispersion characteristics and spectral evolution simulation. a, b** Evolution of the supercontinuum along the propagation distance simulated based on the GNLSE for the (**a**) TE mode and (**b**) TM mode. The TE mode evolution clearly illustrates the soliton fission point, followed by dispersive wave generation (towards shorter wavelengths) and soliton self-frequency shift (towards longer wavelengths). **c** Calculated dispersion curves for the fundamental TE (black line) and TM (red line) modes.

The numerical simulation results (Fig S2a for TE mode and Fig S2b for TM mode) clearly reveal the fundamental differences in their ultrafast nonlinear dynamic evolution. The significant disparity in dispersion characteristics ultimately leads to distinct spectral evolution outcomes[3].

Initial Broadening Stage: Spectral broadening is primarily driven by self-phase modulation. Due to the smaller effective mode area ($A_{\text{eff}}$, see Table S1) of the TE mode, the accumulation rate of nonlinear phase shift per unit length is slightly higher than that of the TM mode. Consequently, the spectral edges of the TE mode expand faster, exhibiting a more pronounced trend of initial supercontinuum broadening.

Soliton Fission Stage: As the pulse compresses further into the high-order soliton regime, the divergence between the two modes amplifies dramatically. Benefiting from a relatively small and flat dispersion profile, the TE mode undergoes clear soliton fission and forms stable fundamental solitons, laying the foundation for the subsequent efficient Raman self-frequency shift. In contrast, the TM mode is constrained by an extremely large anomalous group velocity dispersion ($\beta_2$), undergoing severe temporal broadening before reaching the fission threshold. The pulse energy is rapidly dispersed, and the peak power drops sharply, making it difficult to maintain a high-intensity localized wave packet, thereby significantly suppressing subsequent soliton dynamics.

Frequency Shift and Radiation Stage: In the TE mode, the fundamental solitons generated by fission achieve efficient redshift driven by the Raman effect, with the final center wavelength exceeding 2000 nm. This is accompanied by significant dispersive wave radiation at shorter wavelengths, with radiation peaks located in the visible region. Conversely, the TM mode exhibits a characteristic "dispersion-dominated energy collapse": the strong dispersion effect rapidly dilutes the pulse peak power to below the nonlinear threshold, prematurely truncating the soliton self-frequency shift process and ultimately resulting in a severely limited overall spectral bandwidth.

In Conclusion: The numerical simulations reproduce the experimentally observed spectral evolution characteristics well in terms of macroscopic trends. It is particularly worth emphasizing that although the TM mode in the experiment indeed suffers from higher interfacial scattering loss (approximately 20% power attenuation), this magnitude of loss is far insufficient to solely account for the drastic collapse of the TM mode's spectral bandwidth. The simulation results compellingly demonstrate that the difference in dispersion dynamics, rather than mere loss limitations, is the core physical mechanism dominating the final spectral evolution path and defining the bandwidth boundaries.

## 3. Waveguide Mode Analysis

We performed numerical simulations of the waveguide modes using the full-vectorial finite-element method implemented in COMSOL Multiphysics[4]. The waveguide geometry was defined to match the experimentally fabricated device ($W$=1.0 µm, $H$=800 nm), employing an asymmetric air-cladding configuration.

**(1) Mode Field Distribution and Interface Effects**

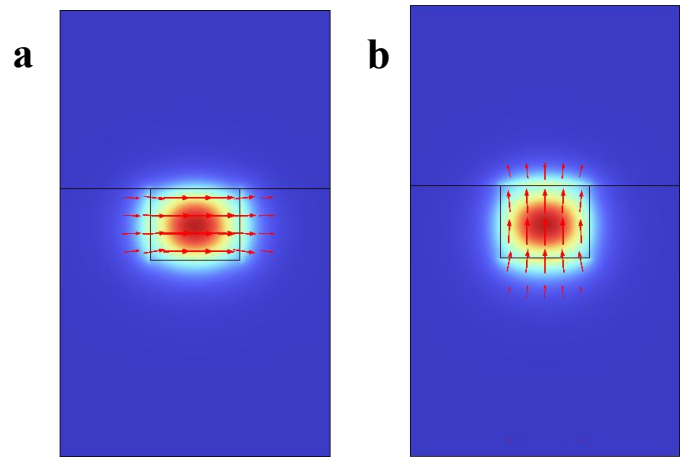

**Fig S3 Simulated mode fields and vector distributions.** Simulated electric field intensity distributions and electric field vector fields (red arrows) for the (**a**) TE fundamental mode and (**b**) TM fundamental mode at a pump wavelength of 1550 nm. The parallel vector field in (**a**) indicates that the TE mode is continuous at the boundary, whereas the vector field perpendicular to the interface in (**b**) exhibits a distinct discontinuity and amplitude jump at the air-core boundary, visually verifying the interfacial field enhancement effect caused by the high refractive index contrast.

Fig S3 illustrates the electric field intensity distributions and electric field vector distributions (red arrows) for the fundamental TE and TM modes at a pump wavelength of 1550 nm. The optical field energy of both modes is tightly confined within the $Ta_2O_5$ core. However, the red vector arrows clearly reveal the fundamental differences between the two at the boundaries.

For the TE mode (Fig S3a), the red vector arrows indicate that its dominant electric field component ($E_x$) is primarily parallel to the upper and lower interfaces of the waveguide. According to electromagnetic boundary conditions, the tangential electric field component remains continuous across the interface; thus, the TE mode exhibits no significant field discontinuity at the boundary. In contrast, for the TM mode (Fig S3b), the vector arrows indicate that its dominant component ($E_y$) is primarily perpendicular to the core-air interface. According to the continuity condition of the normal electric displacement vector ($D_{air} = D_{core}$), and given the substantial refractive index contrast between the core and air ($n_{core} \approx 2.1$, $n_{air} \approx 1.0$), the electric field intensity must undergo an abrupt change at the boundary, satisfying $E_{air} \approx n^2_{core} E_{core} \approx 4.4 E_{core}$. The significant change in the length of the red vector arrows crossing the upper boundary in Fig S3b intuitively demonstrates this field discontinuity and the field enhancement effect on the air side of the interface. According to waveguide sidewall scattering theory[5], scattering loss is proportional to the square of the electric field modulus at the interface. Consequently, the extremely high field intensity at the interface renders the TM mode more sensitive to the surface roughness of the waveguide sidewalls than the TE mode. This provides a quantitative explanation for the slightly higher linear propagation losses observed for the TM mode in experiments.

To further quantify the transmission characteristics of the two modes, we calculated their effective refractive index ($n_{eff}$), effective mode area ($A_{eff}$), and mode confinement factor (Γ), with detailed data listed in Supplementary Table S2. First, the $A_{\text{eff}}$, a critical parameter determining the strength of nonlinear interactions, is given by the formula:

$$A_{\text{eff}} = \frac{\left(\iint |E(x,y)|^2 dxdy\right)^2}{\iint |E(x,y)|^4 dxdy}$$

The calculated results indicate that the $A_{\text{eff}}$ of the TE mode (0.78 μm²) is slightly smaller than that of the TM mode (0.85 μm²). As the nonlinear parameter is inversely proportional to the effective mode area, this implies that the TE mode possesses a higher optical power density and a larger effective nonlinear coefficient. This explains why the TE mode exhibited superior spectral broadening efficiency under equivalent pulse energies in the experiment.

Secondly, the mode confinement factor (Γ) quantifies the proportion of optical power within the waveguide core. The formula is as follows[6]:

$$\Gamma = \frac{\iint_{\text{core}} S_z(x,y) dxdy}{\iint_{\text{total}} S_z(x,y) dxdy}$$

The calculation results demonstrate that both TE and TM modes exhibit extremely high confinement factors (90.9% and 90.7%, respectively), with a negligible difference between them (ΔΓ ≈ 0.2%). This result compellingly rules out "substrate leakage due to weak confinement" as the primary cause for the slightly higher loss of the TM mode. Combined with the mode field analysis discussed previously, we confirm that surface scattering induced by interfacial field enhancement is the dominant mechanism responsible for the slightly higher loss observed in the TM mode compared to the TE mode.

Table S2 Calculated optical characteristics of waveguide modes. Effective refractive index ($n_{eff}$), effective mode area ($A_{eff}$), and mode confinement factor ($\Gamma$) for TE and TM modes at 1550 nm.

| Mode | Effective Index ($n_{eff}$) | Effective Area ($A_{eff}$) | Confinement Factor ($\Gamma$) |
|---|---|---|---|
| TE | 1.86 | 0.78 | 0.909 |
| TM | 1.83 | 0.85 | 0.907 |

## 4. Spectral Broadening and Polarization Characteristics via 60 fs Pulses

To explore the spectral broadening potential of the $Ta_2O_5$ waveguide platform under extreme conditions, we further optimized the pumping conditions in the experiment by compressing the incident pulse duration from 80 fs to 60 fs. This modification effectively boosts the peak power at a fixed pulse energy, thereby enabling a more intense excitation of the nonlinear effects within the waveguide. Specifically, the elevated peak power directly increases the soliton order ($N$), accelerating the dynamics of supercontinuum generation. This rapid evolution mechanism is pivotal; it ensures that the pulse enters the soliton fission stage almost immediately at the waveguide input, thereby efficiently initiating the nonlinear frequency conversion process before propagation losses can significantly deplete the pulse energy.

Fig S4a illustrates the spectral evolution as a function of coupled energy under these pulse conditions. Compared to the 80 fs case, the 60 fs ultrashort pulse significantly shortens the soliton fission length ($L_{fiss}$). This implies that the fundamental solitons generated after fission possess a longer effective propagation distance to undergo Raman self-frequency shifting, driving the red edge of the spectrum to extend beyond 3400 nm.

Ultimately, at a pump energy of approximately 300 pJ, we achieved an ultra-broad supercontinuum spanning nearly three octaves. Notably, even at low energy (92.7 pJ), the spectrum was also broadened to span over two octaves., further confirming the advantage of 60 fs pulses in enhancing energy efficiency.

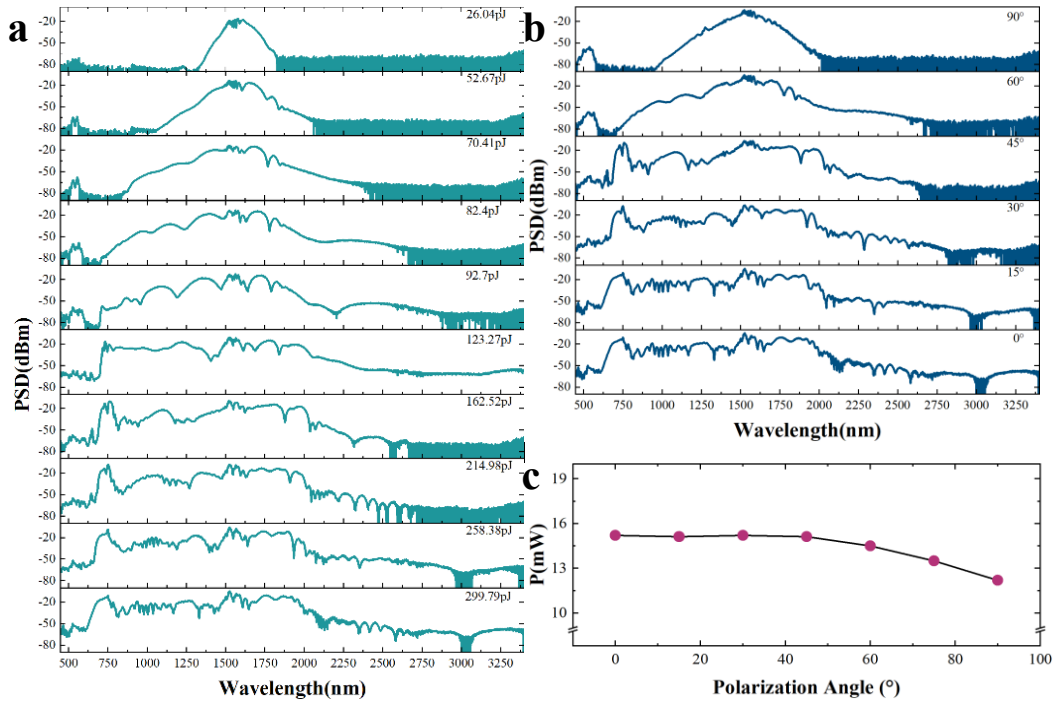

**Figure S4 Extreme spectral broadening characteristics driven by 60 fs pulses. a** Spectral evolution as the coupled pulse energy increases from 26.04 pJ to 299.79 pJ. The input pulse duration is 60 fs with a center wavelength of 1550 nm. **b** Evolution of the supercontinuum as a function of the incident polarization angle (from 0°/TE mode to 90°/TM mode) under maximum pumping conditions. **c** Measured average output power as a function of the polarization angle.

Furthermore, we investigated the polarization dependence under this extreme bandwidth regime (Fig S4b). Experimentally, we observed that as the polarization angle evolved from 0° (TE mode) to 90° (TM mode), the total spectral bandwidth exhibited a distinct shrinking trend. As discussed in Supplementary Note 1, due to the larger dispersion magnitude encountered by the TM mode, nonlinear spectral broadening is subject to stricter physical constraints, leading to the gradual suppression of visible components and a significant redshift of the short-wavelength cutoff edge. Although the overall bandwidth narrows with polarization rotation, it is noteworthy that the device maintains robust performance in the long-wavelength region (Near- to Mid-IR) without spectral disconnection. Simultaneously, the trend in measured output power with varying polarization states provides further direct experimental corroboration of our mode analysis predictions. This result indicates that while the generation of extreme ultra-broadband spectra primarily relies on the TE mode, the waveguide possesses good polarization applicability in the telecommunications and mid-infrared sensing bands (e.g., 1300-2100 nm).

## 5. Interferometric Measurement Setup and Spectral Characterization

To precisely characterize the temporal coherence, axial resolution, and displacement measurement capability of the light source, we constructed an interferometer based on the Michelson architecture. The schematic of the experimental setup is illustrated in Fig S5a.

**(1) Optical System Design**

Given the extremely broad spectral coverage of the supercontinuum source (spanning from the visible to the mid-infrared), conventional lens-based systems would introduce severe chromatic aberration, leading to focal plane shifts and wavefront distortion, which would, in turn, degrade the quality of the interference signal. To address this challenge, we employed off-axis parabolic mirrors in both the sample and reference arms for beam collimation and focusing. This all-reflective design completely eliminates chromatic aberration, ensuring simultaneous and precise focusing of light across different wavelengths, thereby guaranteeing high-contrast interference fringes over the entire ultra-broadband range.

**(2) Data Acquisition and Signal Processing**

Light from the supercontinuum source passes through a broadband 50:50 beam splitter and is split into two arms. The interference signals are recorded using two optical spectrum analyzers (OSAs).

Figs S5c and S5d present the raw spectral interference fringes recorded in the short-wavelength band (1150-1450 nm) and the long-wavelength band (1600-2000 nm), respectively. This directly confirms the excellent coherence of the light source within each sub-band. The point spread functions presented in the main text were obtained by performing k-space resampling and Fast Fourier Transform on these spectral interference fringes[7].

**(3) Measurement Range Verification**

To verify the large-range measurement capability of the system, we varied the optical path difference between the two arms by translating the off-axis parabolic mirror. Fig S5b displays the PSFs ranging from a near position of approximately 6.88 μm to a far position of 4.8 mm. This is consistent with the sensitivity roll-off analysis in the main text, demonstrating the system's capability for millimeter-level deep depth probing. Furthermore, the axial resolution data presented in Figs 3b and 3c of the main text were acquired at the axial position corresponding to the blue curve, i.e., 504μm.

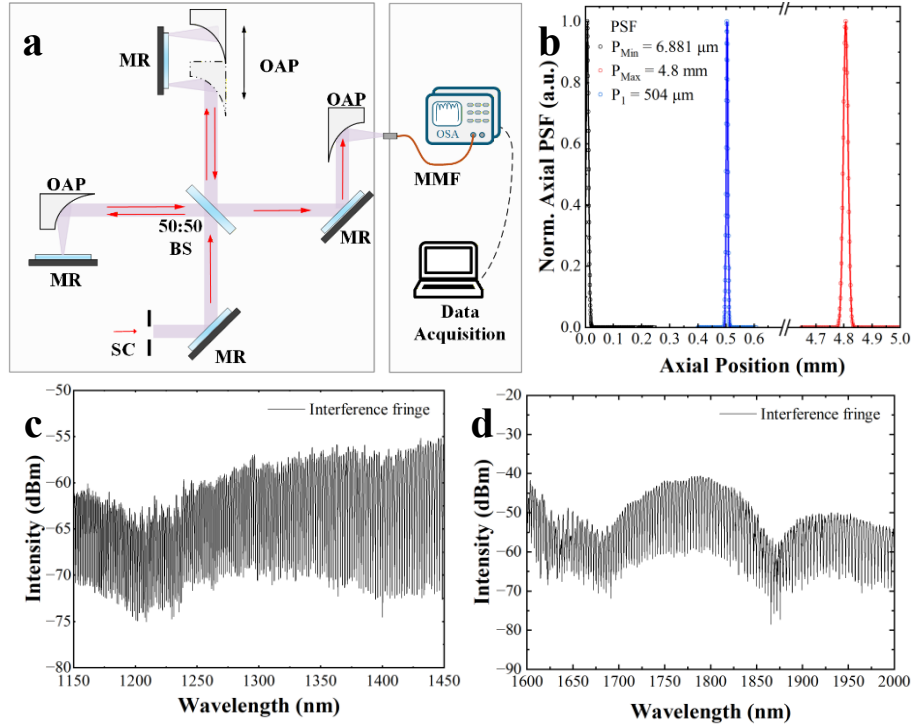

**Fig S5 Experimental setup for interferometric measurement and signal characteristics. a** Schematic diagram of the Michelson interferometer setup used for evaluating the resolution and displacement sensitivity of the light source. **b** Normalized axial point spread functions measured at three representative axial positions: the nearest position ($P_{min} \approx 6.88$ μm), an intermediate position ($P_1 \approx 504$ μm), and a distal position ($P_{max} \approx 4.8$ mm). **c, d** Raw spectral interference fringes recorded by optical spectrum analyzers (OSAs) for the (**c**) short-wavelength band (1150-1450 nm) and (**d**) long-wavelength band (1600-2000 nm), respectively. The high-contrast interference fringes validate the excellent spatiotemporal coherence of the light source.

### (4) Statistical Analysis of Raw Data and Verification of Measurement Limits

We present the complete set of raw acquisition data for the nanometric displacement tests in Fig S6, which clearly illustrates the stepwise variations in measured displacement as a function of the driving pulses applied to the displacement stage.

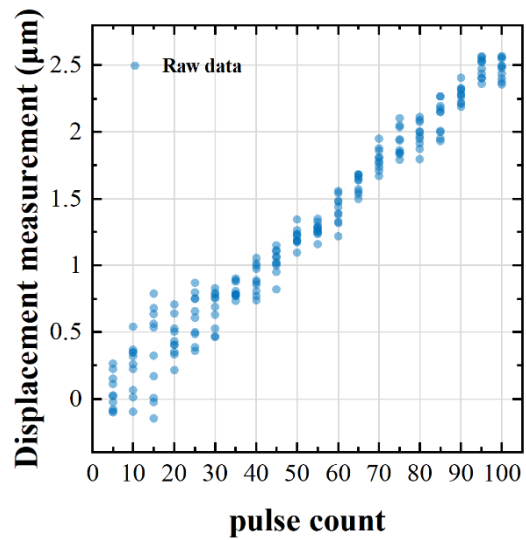

**Fig S6 Scatter plot of raw data for nanometric displacement measurements.** The blue scatter points represent the raw displacement values recorded at different numbers of driving pulses. We performed multiple repeated measurements at each designated step position. These data provide the statistical basis for the error bars plotted in **Fig 3d** of the main text and for the calculation of the minimum resolvable displacement (MRD, 3σ).